\documentclass[prl,twocolumn,showpacs,amsmath,amssymb,superscriptaddress]{revtex4}
\usepackage{graphicx}
\usepackage{hyperref}

\newcommand{\be}{\begin{equation}}
\newcommand{\ee}{\end{equation}}
\newcommand{\bea}{\begin{eqnarray}}
\newcommand{\eea}{\end{eqnarray}}
\newcommand{\p}{\partial}

\renewcommand{\>}{\rangle}
\newcommand{\<}{\langle}

\begin{document}

\title{Orthogonality catastrophe and  shock waves in a non-equilibrium Fermi gas}
\author{E. Bettelheim}
\affiliation{James Frank Institute, University of Chicago, 5640 S.
Ellis Ave. Chicago IL 60637.}
\author{A. G. Abanov}
\affiliation{Department of Physics and Astronomy,
Stony Brook University,  Stony Brook, NY 11794-3800.}
\author{P. Wiegmann}
\affiliation{James Frank Institute, University of Chicago, 5640 S.
Ellis Ave. Chicago IL 60637.} \affiliation{Also at Landau
Institute of Theoretical Physics, Moscow, Russia}

\begin{abstract}

A  semiclassical wave-packet propagating in a dissipationless
Fermi gas inevitably enters  a ``gradient catastrophe'' regime,
where an initially smooth front develops large gradients and
undergoes a dramatic shock wave phenomenon. The non-linear effects
in electronic transport are due to the curvature of the electronic
spectrum at the Fermi surface. They can be probed  by a sudden
switching of a local potential. In equilibrium, this process
produces a large number of particle-hole pairs, a phenomenon
closely related to the Orthogonality Catastrophe. We study a
generalization of this phenomenon to the non-equilibrium regime
and show how the Orthogonality Catastrophe cures the Gradient
Catastrophe, by providing a dispersive regularization mechanism.

\end{abstract}
\date{\today}

\maketitle

%%%%%%%%%%%%%%%%%%%%%%%%%
\paragraph{1. Introduction.}
When a Fermi gas is  perturbed by a sudden switch of a local
potential, it produces soft  particle-hole pairs whose number
grows as $\log (p_F L)$ with the size of the system. This
phenomena is known as Orthogonality Catastrophe
\cite{Anderson:Catastrophe}. It means  that the overlap of the
ground state of the Fermi gas  with a localized potential $|B_a\>$
(a state  emerging as a result of a shake-off) with the
unperturbed ground state $\<0|$ decays with the size of the system
as $\<0|B_a\>\sim (p_FL)^{-a^2}$, where $a=-\delta/\pi$ and
$\delta$ is a scattering phase of  the potential.

The effects of the Orthogonality Catastrophe are observed as the
Fermi Edge Singularity - a power law resonance at the Fermi level
occurring in  transition rates  in x-ray
\cite{NozieresDedominicis} or in tunneling experiments
\cite{Matveev:Larkin}. Recently this phenomenon has been exploited
in a measuring device detecting local charge distribution in
mesoscopic conductors.

The Orthogonality Catastrophe manifests itself differently in
systems out of equilibrium  where energy relaxation is small and
electrons can diffuse out of the system without energy
dissipation. The interest to tunneling out of equilibrium states
is growing, but apart from recent works
\cite{Levitov:Abanin:Fermi:Edge:Noneq,Muzykantskii:DAmbrumenil:Braunecker}
little is known. Perhaps  the reason  is that this problem can not
be approached by methods traditionally used for equilibrium
states.

%%%%%%%%%%%%%%%%%%%%%%%%%
\paragraph{2. Transition rates.}
Consider a non-equilibrium state $\<g|$ initially created in the
Fermi gas, where we assume no interaction and ignore spin. This
state evolves with the Hamiltonian
\begin{equation}
H_0 = \sum_p  \frac{p^2}{2m} \psi^\dagger_p \psi_p
\end{equation}
as $\<g(t)|=\<g|e^{-i H_0 t}$. At time $t$ we probe the state at
the point $x$ by a sudden switch of a local potential $U(x)$.

In this letter we ask the following question. What is the
probability to find the system in the ground  state  $|B_a(x)\>$
of the the new, perturbed Hamiltonian $H = H_0 +U$ at time $t$. In
other words, we are looking for a space-time dependence of the
transition amplitude
\begin{equation} \label{tau}
\<g|e^{-iH_0 t}|B_a(x)\>.\quad
\end{equation}
Such transition rates can be measured in transport experiments
similar to those of Ref.\cite{Cobden}. There a  quantum dot with a
resonant level  was brought into a proximity of a  Fermi gas. When
an electron tunnels out of the Fermi gas to the dot, the dot
produces a potential seen by the Fermi gas a sudden shake-off. If
the level suddenly  becomes unoccupied the transition amplitude
reads
\begin{equation} \label{tau1}
\<g|e^{-iH_0 t}\psi(x)|B_{a}'(x)\>,
\end{equation}
where $|B_{a}'(x)\>$ is a ground state of the perturbed system with one extra particle.

Other measurable quantities, e.g.,  tunneling current
\cite{NozieresDedominicis,Matveev:Larkin} or generating functions
of quantum noise , involving projections of evolving states
$\<g(t)|$ onto states of the perturbed gas other than the ground
state, can also be computed using the methods developed below.

Assume that the probing potential causes no backscattering and is
well localized, so that a scattering phase $\delta$ can be treated
as a constant in the range of momenta of the wave-packet $|g\>$.
Then a perturbed Hamiltonian is $H=e^{a\varphi(x)}H_0
e^{-a\varphi(x)}$ \cite{SchotteSchotte} and its ground state is
\begin{equation}
\nonumber |B_a\> = e^{a\varphi}|0\>=(p_F L)^{-a^2}\! : \!
e^{a\varphi}\! :\! |0 \>, \quad a = - \delta/\pi,
\end{equation}
where the normal ordering separates the equilibrium part of the
Orthogonality Catastrophe. Here $\varphi(x) =  2 \pi i \int^x
\rho(x') dx' $ is an antihermitian chiral Bose field and
$\rho(x)=\psi^\dag(x)\psi(x)$ is the fermionic density. Similarly,
$\psi(x)|B'_a(x)\>\sim e^{\varphi(x)}|B_a(x)\>\sim
e^{(a+1)\varphi(x)}|0\>$, where we used a ``bosonization'' formula
and set $-1< a\leq 0.$

Summing up, we study space-time dependence of the transition rates
\begin{equation}\label{e}
\tau_a \!=\!\< g(t)|\!:\! e^{a\varphi(x)}\!:\!|0 \>,\;\;
\tau_{a+1} \!=\! \< g(t)|\!:\! e^{(a+1)\varphi(x)}\!:\!|0 \>,
\end{equation}
where we dropped  space-time independent factors representing the
equilibrium part of Orthogonality Catastrophe. We denote
logarithmic derivatives as
\begin{equation}\label{u}
u = i  \frac{\hbar}{m} \p_x \log \frac{ \tau_a}{ \tau_{a+1}}  , \quad
\tilde u = \frac{\hbar}{m} \p_x \log{ \tau_a } { \tau_{a+1} }.
\end{equation}

We will see that the rates (\ref{e}) undergo complicated
dynamics, experiencing a shock-wave and a subsequent set of
oscillations filling a growing spatial region. In fact, the shock
wave occurs even at $a=0$ (or integer), without the Orthogonality
Catastrophe. However, its physics and the scale
of oscillations are essentially different \cite{Damski:Shock:Waves}.

\begin{figure}[t!!!]
\includegraphics[width=6cm]{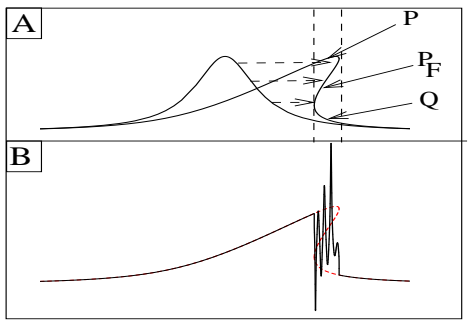}
\caption{\label{thefigure} [Color Online] A:  A shock-wave
solution of the Riemann equation (\protect\ref{H}). The dashed
arrows indicate the velocity of the front. The vertical dashed
lines are trailing and leading edges.  The solid arrows show the
relations between branches of the multi-valued solution of the
Riemann equation and Whitham modulated particle $P$, hole $Q$ and
Fermi $P_F$ momenta. $B$: Oscillations obtained by the Whitham
method, the dashed red line shows the unphysical part of the
Riemann solution. }
\end{figure}

%%%%%%%%%%%%%%%%%%%%%%%%
\paragraph{3. Semiclassical and coherent states.}
We will especially be interested in semiclassical wave-packets,
i.e., states  whose Wigner function $ W(x,p) = \< g | e^{
\frac{i}{\hbar} ( P x + X p )} | g \>$, where $P$ and $X$ are
momentum and coordinate operators, initially localized in the area
of the phase space $\Delta x \gg \hbar p_F^{-1}$ and $\Delta p= |
p - p_F | \ll p_F$. This packet carries a large number of
particles $N = 2 \pi \hbar \Delta x \Delta p \gg 1$. We may choose
such a state to be coherent, i.e., given by $\<g| = \< 0 | e^{
\sum_{pq} A_{pq} \psi^\dag_p \psi_q } $. This state corresponds to
a smooth localized bump of electronic density as on
Fig.\ref{thefigure} and can be created by the action of a
classical instrument. We also assume that the distance $x$ between
the initial origin of the wave-packet and a point of the
measurement is large $x\gg \Delta x$.

%%%%%%%%%%%%%%%%%%%%%%%
\paragraph{4. Hydrodynamic interpretation and the role of Orthogonality Catastrophe.}
In the semiclassical approximation the amplitudes (\ref{e})
acquire a useful hydrodynamic interpretation. Let us assume that
$A_{p,p+k}=A_{k}$ depends only on the momentum change, $k$,  and
write the initial state as $\<g | = \<0 | e^{ \int V_+(x)
\varphi(x) dx }$, where $ V_+ (x) = \sum_{k>0} A_{k}
\frac{e^{ikx}} { 2 \pi  k}$, where $V_+(x)$ is an analytic
function in the upper half-plane of $x$. In this case the density
of the classical wave packet $  \< g | \rho(x) | g \>=-\frac{1}{2
\pi } {\rm Im} V'_+$. On the other hand the initial values of the
amplitudes are $\tau_a(x)=e^{ a V_+( x ) }$, and, therefore,
initially $  \frac{2\pi\hbar}{m}   \< g | \rho(x) | g \> = {\rm
Re}\, u(x). $

In the course of the evolution  the above relation between the
density and the amplitudes is destroyed. However,  in the
semiclassical limit and   if $a\neq \mbox{integer}$ the rates
(\ref{e}) still contain all the hydrodynamic information.

%%%%%%%%%%%%%%%%%%%%%%
\paragraph{5. Dispersion of the electronic spectrum
and non-linearity of the waves.}
It is commonly assumed that the linearization of the electronic
spectrum at the Fermi surface $ H - E_F \approx \sum_p v_F( p -
p_F) \psi^\dag_p \psi_p$ captures the physics of the Orthogonality
Catastrophe. If this were so, the time dependence of
transition rates would be no different than its space dependence. The
state $\< g(t) | = \< g | e^{-\frac{i}{\hbar} v_F P t}$ simply
translates the point of measurement: $\tau(x,t)=\tau(x-v_Ft)$
without any interesting dynamics.

However, the approximation of linear spectrum is valid only for
some time $t\ll t_c$. It {\it inevitably breaks down at larger
time}. The physics is simple: electrons in the denser part of the
packet  at the top of the bright side of the bump have  higher
momenta $\delta p=p-p_F= \hbar\delta\rho$ and, therefore,  move
with higher velocities $ v - v_F = \frac{1}{2} \hbar E''(p_F)
\delta p$ than particles in front of them. Here $\frac{1}{2} \hbar
E''(p_F)=\frac{\hbar}{2m}$ is a curvature of the spectrum at the
Fermi point \cite{Glazman}. As a result, the wavefront steepens
and eventually overturns (Fig.~\ref{thefigure}). This is the shock
wave we study using amplitudes (\ref{e}) as a ``measurement''. The
results of the ``measurement'' depend on $a$ and are especially
sensitive to whether $a$ is an integer or not.

The critical time of entering into the shock wave regime is about
the time wave packet crosses the distance equal to the size of its
front $t_c\sim\frac{m\Delta x}{\Delta p}$. We assume that $t_{c}$
is smaller than the ballistic time, so that dissipative effects in
real systems do not have time to dissipate the shock.

%%%%%%%%%%%%%%%%%%%%%%%%%%
\paragraph{6. MKP equation of the soliton theory.}
Non-linear aspects of electron dynamics, can not be analyzed by
elementary means. We have derived a fundamental equation which
determines  both  rates (\ref{e}). It is  the {\it modified
Kadomtsev-Petviashvili } equation (or MKP) - a known equation in
soliton theory \cite{paperI}. Its bilinear form reads
\begin{equation}\label{MKP}
(iD_t - \frac{\hbar}{2m} D_x^2) \tau_{a}\cdot \tau_{a+1}=0,
\end{equation}
where  $D_x f\cdot g=f'g-fg'$ is the Hirota derivative. In fact,
this equation  holds for a more general class of matrix elements
$\<g|e^{a\varphi(t)}|h\>$, where $|h\>$ is any coherent state. We
sketch the proof of the MKP at the end of this letter.

Solutions of the MKP must be sought in the class of
functions analytical in the upper half  of the complex plane
$x$. These are the properties of the matrix elements with respect
to the Fermi vacuum - momenta of all  excitations  exceed the
Fermi momentum. Analytical conditions
are important. In particular they exclude soliton solutions of the
MKP equation.

In terms of (\ref{u}) we have another form of the MKP:
\begin{equation}\label{uMKP}
\dot u = u\p_x u+\frac{\hbar}{2m} \p_x^2 \tilde u.
\end{equation}
At $a=0$, $\tau_0=1$, and  $u=\tilde u$. In this case a non-linear MKP
equation becomes a linear Schr\"{o}dinger equation for $\tau_1 =
e^{\frac{im}{\hbar} \int^x udx}$. Similarly, (\ref{uMKP}) becomes a
``complex Burgers equation''. Even in this simple case the dynamics
of a semiclassical  wave packet is not simple, but it can be
obtained by elementary  means studying the semiclassical limit
of the solution of Shr\"{o}dinger equation in the space of
analytical functions in the upper half-plane.

Analysis at $a\neq 0$ requires  methods of soliton theory.

%%%%%%%%%%%%%%%%%%%%%%%%%
\paragraph {7. Multi-phase  solution of the MKP equation.}\cite{paperI}.
Assume that the initial state consists of a finite number
of particles with momenta $p_i>p_F$ and holes with momenta
$q_i<p_F$, such that $\<g|=\<0|e^{\sum_{i\leq N}
A_{p_iq_i}\psi^\dag_{p_i}\psi_{q_i}}$. Then \\
$\tau_a = e^{\frac{i}{\hbar} a\theta_F} \det_{i,j} (\delta_{ij} +K_a(p_i,q_j))$,
where
\begin{equation}\label{K}
    K_{a}(p_i,q_j) = \frac{\sin(\pi a)}{\pi} A_{p_iq_i}
    \left( \frac{ p_i- p_F }{ p_F - q_i } \right)^a \frac{ e^{\frac{i} {\hbar}
\theta_{i}( x , t )}}{ p_i - q_j},
\end{equation}
and $ \theta ( p_i , q_i ) = ( p_i - q_i ) x - \frac{1}{2m}( p_i^2
- q_i^2 ) t$, $\theta_F = p_F x - E_Ft$. A formal solution for
generic initial data is given by the determinant of the Fredholm
operator $\mathbf{1+K}$.

This result can be obtained directly from the definition of the
matrix elements. One identifies the kernel  $K_a(p,q)$ as the
particle-hole  amplitude $K_a(p,q) = \<g|p q\> \< p q | e^{ i P x
- i H_0 t } | p q \> \< p q | B_a \> $, and $A_{ p q } =\< g | p q \> $.
The matrix element   $ \<p q | B_a \>$ is the overlap between
a particle-hole pair and the ground state of the perturbed Fermi gas
computed in \cite{paperI} $ \<p q | B_a \>= \left( \frac{ p - p_F }{ p_F -q
} \right)^a \frac{ \sin( \pi a)}{ \pi ( p - q ) } $ for $ p \neq q
$. Its singularity at the Fermi energy is a signature of the Orthogonality
Catastrophe.

In particular, the 1-phase solution is ($p>p_F>q$)
\begin{equation}
 \label{1phase}
    \tau_a\! =e^{\frac{i}{\hbar} a\theta_F} \left[1+A_{pq}\!\frac{\sin(\pi a)} {\pi}
 \!\left( \frac{ p - p_F}{
p_F - q } \right)^a\! \frac{ e^{\frac{i}{\hbar} \theta(p,q)}}{ p - q
}\right].
\end{equation}

%%%%%%%%%%%%%%%%%%%%%%%%%
\paragraph{8. Quantum Riemann equation.}
In order to understand the MKP equation we recount the formulation
of 1D Fermi gas as quantum hydrodynamics, also known as
(non-linear) bosonization, or collective field theory
\cite{Jevicki:1991yi}. The quantum equation of motion  of the chiral
Fermi gas can be  cast entirely in terms of the
density operator
\begin{equation} \label{H}
\p_t {\rm u} = {\rm u} \partial_x {\rm u}  ,\quad {\rm u} =
\frac{2 \pi\hbar}{m} \rho.
\end{equation}
This is the quantum Riemann equation. This equation holds on
coherent states generated by the density operator. Its proof
consists of a check that its l.h.s. commutes with all density
modes $\rho_k = \int e^{-i kx } \rho(x) dx$. In their turn, the
modes form the current algebra

%%%%%%%%%%%%%%%%%%%%%%%%%
\paragraph{9. Classical Riemann equation and Shock waves.}
We note that the first two terms in the classical MKP equation
(\ref{uMKP}) are exactly the same as in the semiclassical version
of the  Riemann equation (\ref{H}). This is no surprise, since
they are uniquely determined by  the Galilean invariance. One can
neglect the third term in (\ref{uMKP}), or the quantum correction
in (\ref{H}) if the gradients are small. They, indeed,  are
assumed to be small initially. Also under a semiclassical
condition one can neglect an imaginary part of $u$. Then eq.
(\ref{H}) becomes  the Riemann equation of compressible
hydrodynamics \cite{Landau:Book:Section:101}.

Riemann's equation leads to shock waves: the velocity of a point
with height $u(x)$ is $u(x)$ itself - higher parts of the front
move with higher velocities. The  bright side ($u\p_xu<0)$ of any
smooth initial data  gets steeper, and eventually achieves an
infinite slope $\p_x u(x_c,t_c)=\infty$ at some finite time
$t=t_c$ - a shock wave.

After this moment the Riemann equation has at least three real
solutions (Fig.\ref{thefigure}) confined between
$x_-(t)$ - the trailing edge, and $x_+(t)$ - the leading edge.
They, and the critical point $t_c$ can be easily found from the
implicit solution due to Riemann $u(x,t)=f(x-u(x,t)\cdot t)$,
where $f(x)=u(x,0)$ is the initial profile. For a typical wave
packet with a height $\Delta p/m$ and
width $\Delta x$ a critical time is of the order of
$t_c\sim m\Delta x/\Delta p$.
%$t_c\sim(m/\hbar)(\Delta x)^2/ n$.
The leading edge (in the Galilean frame
moving with velocity $v_F$) moves with velocity $\Delta p/m$: $x_+\sim
(\Delta p/m) t$. If $f(x)\sim x^{-n}$ at $x\gg \Delta x$ the trailing edge
delays, progressing as $x_-\sim \Delta x(t/t_{c})^{1/(n+1)}$.

Let us label the solutions of the Riemann equation at $t\!\!>\!\!t_c$:\!
$u^{(0)}$ is a single-valued solution outside the shock wave
interval,  $u^{(1)}\!\!>\!\!u^{(2)}\!>\! u^{(3)}$ are three ordered solutions
in the shock wave interval $x_-(t)\!<\!x\!<\!x_+(t)$. The branch $u^{(1)}$
smoothly merges with $u^{(0)}$ at the trailing edge, while
$u^{(3)}$ smoothly merges with $u^{(0)}$ at the leading edge.

%%%%%%%%%%%%%%%%%%%%%%%%%%%
\paragraph {10. Dispersive regularization and the role of Orthogonality Catastrophe.}
Obviously, the approximation leading to the Riemann equation fails
when gradients are large. Then, the neglected gradient terms become
important. They regularize the ``gradient catastrophe''. The
regularization and the subsequent physics of the shock wave
is very different for an integer $a=0$, and in the case of
the Orthogonality Catastrophe, where  $a$ is
irrational (a rational $a$ involves some  additional structures).

If $a=0$ the solution of the Schr\"{o}dinger equation at
$t\!>\!t_c$  is well approximated by $u^{(0)}$ at $x\!<\!x_-(t)$
with an abrupt fall to $u^{(3)}$ at $x\!>\!x_-(t)$. In the case of
the Orthogonality Catastrophe ($a\neq$ integer), we  face
complexity of the nonlinear equation (\ref{uMKP}). In this case
the entire  interval\!  $x_-\!\!<\!\!x\!<\!x_+$ is filled by
oscillations (Fig\! \ref{thefigure}).

%%%%%%%%%%%%%%%%%%%%%%%%%%%
\paragraph{11. Whitham modulation.}
Despite the integrability of the MKP equation, its solution in the
form of the Fredholm determinant, with the kernel (\ref{K}), is
rather complicated and the initial value problem is generically
difficult to solve. Fortunately, a powerful approximate method to
describe the shock waves has been developed in a seminal paper
\cite{Gurevich:Pitavsk}. The method suggests to glue a solution of
the Riemann equation which is valid for $x<x_-(t)$ and $x>x_+(t)$,
to a periodic solution. In the first approximation the 1-phase
solution (\ref{1phase}) can be used. The amplitude and the period
of the  wave have to be modulated in order to match very different
values of the front  at the trailing $x_-(t)$ and the leading
$x_+(t)$ edges of the shock.

Modulated non-linear waves are the subject of the Whitham theory
\cite{Whitham:book}. The latter states that modulated waves  have
the form of a multiphase solution (\ref{K}), which
moduli  and phases  are smooth functions of space-time \\
$p , q , p_F , \theta , \theta_F \to P(x,t),
Q(x,t),P_F(x,t),\Theta(x,t),\Theta_F(x,t)$.

In our case the moduli have a clear physical interpretation. They
are momenta of soft particle-hole pairs produced at the  Fermi
level, which is also changing in space-time. The  phases in
(\ref{K}) obey the Whitham equations:
$$
\dot\Theta\!=\!E(P)\!\!-\!\!E(Q),
 \p_x\Theta\!=\!P\!\!-\!\!Q,\, \dot\Theta_F\!\!=\!\!E(P_F),
 \p_x\Theta_F\!\!=\!\!P_F,
$$
where $E(P)=\frac{P^2}{2m}$ is  a modulated energy.  The Whitham
equations for the moduli are determined by Galilean invariance.
They are again Riemann eqs. \cite{Krichever:BO:Whitham}
\begin{equation}\label{W}
 \dot{P} + \p_x E(P)=\dot{Q} +\p_x E(Q) =
\dot{P}_F+\p_x (P_F)=0.
\end{equation}

The initial data for the Whitham  equations are chosen so that
the 1-phase oscillatory solution (\ref{1phase}) is glued to a
(non-oscillatory)  solution of the Riemann equation $u^{(0)}$ at
the leading and the trailing edges.

Assuming that $A_{pq}$ is smooth, we notice that the 1-phase
solution stops to oscillates when a hole is absorbed at the Fermi
level at the trailing edge, and when a particle is created  at the
Fermi level at the leading edge $ Q(x_-)=P_F(x_-),\quad
P(x_+)=P_F(x_+)$. This yields that at the trailing edge $u(x_-) =
(i/ m) \p_x( \Theta + \Theta_F )$. According to the Whitham
equation $i\p_x(\Theta + \Theta_F)=P-Q+P_F$. which is  just $P$ at
$x=x_-$. Therefore $u^{(1)}=P$ are the boundary data for the
Whitham equation for $P$ at the trailing edge. Since $u^{(1)}$ is
a solution of the Riemann equation $P= u^{(1)}$ holds in the
entire oscillatory interval.

At the  leading edge, $u =(i/ m) \p_x( \Theta_F - \Theta ) =
(1/m)( P_F + Q - P) = Q/m$. Therefore, $u^{(3)}(x_+)=Q(x_+)$ is a
boundary data and also a solution of the Whitham equation for $Q$.
In a similar fashion one concludes that the modulated Fermi
momentum $P_{F}$ is given by the branch $u^{(2)}$ of the Riemann
equation. Summing up
$$  x_-(t)\! <\! x \!<\! x_+(t) : \; P = u^{(1)} < P_F = u^{(2)} <
Q = u^{(3)}. $$ Being substituted into the 1-phase solution
(\ref{1phase}) these formulas give an explicit (approximate)
solution of the MKP  in the oscillatory region shown in
Fig.~\ref{thefigure}. \footnote{The multi-valued real solution of
the classical Riemann equation can be interpreted as a boundary of
the phase space  filled by fermions. From this point of view the
oscillations may be linked to the momentum transfer between
different Fermi points in the shock wave region.}

%%%%%%%%%%%%%%%%%%%%
\paragraph{12. Derivation of the MKP equation.}
We sketch the derivation of the MKP  for the
amplitudes $\tau_a=\<g|e^{a\varphi(x,t)}|h\>$ where  $|h\>$ is a generic
coherent state (we focused on  $|h\>=|0\>$ in the paper). For a
more detailed discussion of the relation between the dynamics  of the
Fermi gas and soliton theory see \cite{paperI}.

First, with the help the
quantum Riemann equation (\ref{H}), we compute the action of the Schr\"{o}dinger operator
\begin{align}\label{11}
& ( i\partial_t + \frac{\hbar}{2m} \partial_x^2  ) \!:\!e^{ a \varphi}\!: =
a(a+1)\!:\! e^{ a     \varphi} T\!:,\! \nonumber \\
& (-i\partial_t + \frac{\hbar}{2m} \partial_x^2  ) \!:\!e^{ (a+1)
\varphi}\!: = a(a+1)\!:\! e^{ (a+1) \varphi} \bar{T} \!:,
\end{align}
where $T=:\varphi'^2: -\varphi''$ and  $\bar{T}=:\varphi'^2:
+\varphi''$ are holomorphic (antiholomorphic) components of the
stress-energy tensor of a chiral Bose field.

Using these formulas we write the  eq. (\ref{MKP})  in the form
\begin{equation}\nonumber
 \frac{\< g|\!T e^{a\varphi}\! | h\>}{ \< g|
e^{a\varphi}| h \>} + \frac{\< g|\!\bar{T}
e^{(a+1)\varphi}\! | h\>}{ \< g| e^{(a+1)\varphi}
| h \>} =
 2 \frac{\< g|\! J e^{a \varphi} \! | h \>}{\< g |
e^{a \varphi} | h \>}  \frac{\< g|\! J e^{(a+1) \varphi}
\! | h \>}{\< g |  e^{(a+1) \varphi} | h \>}
\end{equation}
where $J=\p_x\varphi=2\pi i\rho$ is the current of the Bose field,
and the expressions are understood to be normal ordered.

In terms of fermions $T(x) \sim : \psi^\dag(x) \p_x^2 \psi(x):$,
and using the bosonization formula $:e^{ a \varphi}: \sim
:e^{(a+1) \varphi} \psi^\dagger (x) \psi( i \infty) : $ we rewrite
the numerator of the first term in the l.h.s. as a four fermion
insertion:
\begin{align}
\lim_{z,y\to x}\p^2_{y}
 \<g| \!:\!
\psi^\dagger(z) \psi^ \dagger (x) \psi(y) \psi (i\infty)
 e^{a \varphi(x)   } \!:\! | h \>. \nonumber
\end{align}
One may now apply Wick's theorem, carefully taking care of normal
ordering,  to write this in terms of matrix elements with two
fermion insertions. Then, writing the fermions in terms of the
Bose field and taking the limit one proves the MKP equation
(\ref{MKP}).

%%%%%%%%%%%%%%%%%%%%%%%
\paragraph {13. Summary.}
The  density of the semiclassical wave packet
measured by a sudden switching of the potential initially behaves
according to the classical Riemann equation, i.e., similar to a
propagating disturbance in a classical compressible
liquid. At the time $t_c$ the wave packet enters a shock wave
regime. It collapses emanating particle-hole pairs resulting in
modulated oscillations. The wave vector
%and the frequency
of oscillations is of the order of $\Delta p$ --
%and $(\Delta p)^2/2m$ respectively,
 the  ``height'' of the initial wave packet. It is much smaller than the Fermi scale $p_F$. The oscillations occupy an interval whose leading edge
 propagates with the velocity exceeding the Fermi velocity by
$\Delta p/m$.
The oscillations are  a distinct  signature of the Orthogonality Catastrophe.
An observation of quantum shock waves, say, on the edge of Integer Quantum
Hall Effect would be yet another manifestation of the quantum coherence.
%\footnote{It is interesting to compare these results
%to  an evolution of a wave-packet
%in an interacting electronic liquid - Calogero model, studied in our recent
% paper \cite{paper2}.
%Accidentally, both problems are described by the same MKP
%equation, but require different solutions
%corresponding to a very  different
%physics.
% In the non-interacting case discussed here, the shock wave   ends up
%with dispersing oscillations. Contrary, in the
%interacting case a shock wave produces
% non-dispersing solitons.}.

\paragraph{Acknowledgment.} We have  benefited from discussions
with   I. Krichever and L. Levitov. P.W.  and E.B. were supported
by the NSF MRSEC Program under DMR-0213745 and NSF DMR-0220198.
E.B. was also supported by BSF  2004128.  The work of A.G.A. was
supported by the NSF under the grant DMR-0348358.

\end{document}